\documentclass{amsart}
\usepackage{amsthm}
\usepackage{amsmath}
\usepackage{amsfonts}
\usepackage{amssymb}
\usepackage{mathrsfs}
 \theoremstyle{definition}
 
 \theoremstyle{remark}

 \numberwithin{equation}{section}

\def\N{{\mathbb N}}
\def\R{{\mathbb R}}

\def\Q{{\mathbb Q}}

\def\Z{{\mathbb Z}}

\def\T{{\mathbb T}}
\def\PP{\mathcal P}
\def\NN{\mathcal N}
\def\P{\mathcal P}

\def\eps{\varepsilon}
\def\rho{\varrho}
\def\theta{\vartheta}
\def\Cci#1{C_c^\infty(#1)}

\def\dx{\,\text{d}x}
\def\d{\,\text{\rm d}}
\def\sigmaess{\sigma_{\text{\rm ess}}}
\def\sigmadisc{\sigma_{\text{\rm disc}}}
\def\supp{\text{supp }}

\def\norm#1{\left|\!\left|{#1}\right|\!\right|}

\def\phi{\varphi}
\def\theta{\vartheta}

\begin{document}
%
%
%
%
%
%
%
%
%
\title[Dislocation and rotation problems]
 {Dislocation problems for periodic Schr\"odinger operators and mathematical aspects of small angle grain boundaries}
\author[R. Hempel]{Rainer Hempel}

\address{%
Institute for Computational Mathematics,
Technische Universit\"at Braunschweig,
Pockelsstra{\ss}e 14,
38106 Braunschweig, Germany}

\email{r.hempel@tu-bs.de}

\author[M. Kohlmann]{Martin Kohlmann}
\address{Institute for Applied Mathematics,
Leibniz Universit\"at Hannover,
Welfengarten 1,
30167 Hannover, Germany}
\email{kohlmann@ifam.uni-hannover.de}


\date{\today}
\dedicatory{}

\begin{abstract}
We discuss two types of defects in two-dimensional lattices, namely
(1) translational dislocations and (2) defects produced by a rotation of
the lattice in a half-space.

 For Lipschitz-continuous and $\Z^2$-periodic potentials,
we first show that translational dislocations produce
 spectrum inside the gaps of the periodic problem;
 we also give estimates for the (integrated) density of
the associated surface states. We then study lattices with a small angle
defect where we find  that the gaps of the periodic problem fill
with spectrum as the defect angle goes to zero. To introduce our
methods, we begin with the study of dislocation problems on the
real line and on an infinite strip. Finally, we consider
examples of muffin tin type. Our overview refers to results
 in [HK1, HK2].
\end{abstract}

\maketitle
%
\section{Introduction}
%
%
In solid state physics, pure matter in a crystallized form is
usually described by a periodic Schr\"odinger operator $-\Delta + V(x)$
in $\R^3$,  where the potential $V$ is a periodic function.
 In reality, however, crystals are not perfectly periodic since
the periodic pattern of atomic arrangement is disturbed by
various types of crystal defects, most notably:

--- point defects where single atoms are removed (vacancies)
   or replaced by foreign atoms (impurities),

--- large scale defects that produce a surface at
which two portions of the lattice (or two different half-lattices)
face each other (line defects, grain boundaries).

For the modeling of point defects, random Schr\"odinger operators
are the appropriate setting (cf., e.g., [PF] or [V]).  Here we present a deterministic approach
to some two-dimensional models with defects from the second class.

Let $V\colon\R^2\to\R$ be (bounded and) periodic with respect to the lattice
$\Z^2$ and consider the family of potentials
$$
W_t(x,y):=\left\{
\begin{array}{ll}
 V(x,y), & \text{$x \ge 0$,} \\
 V(x+t,y), & \text{$x < 0$},
\end{array}\right. \qquad t \in [0,1].
\eqno{(1.1)}
$$
We then let $D_t:=-\Delta+W_t$ denote the associated (self-adjoint)
Schr\"odinger operators, acting in $L_2(\R^2)$. The operators $D_t$ are
the Hamiltonians for a two-dimensional lattice where the potential
equals the $\Z^2$-periodic function $V$ on $\{x\geq 0\}$ and a
shifted copy of $V$ for $\{x<0\}$, i.e., we study a {\it dislocation
problem}. We call $W_t$ the {\it dislocation potential}, $t$ the
{\it dislocation parameter} and $D_t$ the {\it dislocation
operators}. The spectrum of $D_0=D_1$ is purely absolutely
continuous and has a band-gap structure,
$$
\sigma(D_0)=\sigmaess(D_0)=\cup_{k=1}^{\infty}[a_k,b_k],\quad
a_k<b_k\leq a_{k+1}.\eqno{(1.2)}
$$
The spectral gaps $(b_k,a_{k+1})$ are denoted as $\Gamma_k$. For
simplicity, we will sometimes write $a$ and $b$ for the edges of
a given $\Gamma_k$ with  $\Gamma_k\neq\emptyset$. We shall say that
a gap $\Gamma_k = (a,b)$ is \emph{non-trivial} if $a < b$ and
$a$ is above the infimum of the essential spectrum of the given
self-adjoint operator.

%
We will show that the operators $D_t$ possess \emph{surface states} (i.e., spectrum
produced by the interface) in the
gaps $\Gamma_k$ of $D_0$, for suitable values of $t\in(0,1)$.
More strongly, we have a positivity result for the (integrated) density of
the surface states associated with the above spectrum in
the gaps. Here we first have to choose an appropriate scaling which
permits to distinguish the bulk from the surface density of states.
To this end, we consider the operators $-\Delta + W_t$ on squares $Q_n
= (-n,n)^2$  with Dirichlet boundary conditions, for $n$ large,
count the  number of eigenvalues inside a compact subset of a
non-degenerate spectral gap of $D_0$ and scale with $n^{-2}$ for
the bulk and with $n^{-1}$ for the surface states. Taking the limit
$n\to\infty$ (which exists as explained in [KS, EKSchrS]), we
obtain the (integrated) density of states measures $\rho_{\text{bulk}}(D_t,I)$
for the bulk and $\rho_{\text{surf}}(D_t,J)$
for the surface states of this model; here $I \subset \R$ and $J
\subset \R\backslash\sigma(D_0)$ are open intervals and $\overline
J \subset \R\backslash\sigma(D_0)$. (The fact that an integrated surface
density of states exists does not necessarily mean it is
non-zero and there are only rare examples where we know
$\rho_{\text{surf}}$ to be non-trivial.) Our first main result can be
described as follows:
\vskip1em

{\bf 1.1.~Theorem.} {\it If $(a,b)$ is a non-trivial spectral
gap of the periodic operator $-\Delta + V$, acting in $L_2(\R^2)$ with $V$ Lipschitz-continuous,
then for any interval  $(\alpha,\beta)$ with $a < \alpha < \beta < b$
there is a $t \in (0,1)$ such that} $\rho_{\text{surf}}(D_t,
(\alpha,\beta)) > 0$.

\vskip1em
We also explain how to obtain upper bounds (as in [HK2]) for the surface density
of states. In Section 2, we will outline a proof of Theorem
1.1 starting from dislocation problems on $\R$ and on the strip
$\Sigma:=\R\times[0,1]$.  The one-dimensional
 dislocation problem has been studied extensively by Korotyaev
 [K1, K2], and we use the 1D model mainly for testing our methods in
 the simplest possible case.


The techniques and results connected with Theorem 1.1 are
mainly presented as a preparation for the study of rotational defects
where we consider the potential
$$
V_\theta(x,y):=\left\{
\begin{array}{ll}
V(x,y), & \text{$x \ge 0$,} \\
V(M_{-\theta}(x,y)), & \text{$x < 0$,}
\end{array}
\right.
\eqno{(1.3)}
$$
where $M_{\theta} \in \R^{2 \times 2}$ is the usual orthogonal matrix
associated with rotation through the angle $\theta$.
The self-adjoint operators $R_\theta:=-\Delta+V_\theta$ in $L_2(\R^2)$ are the
Hamiltonians for two half-lattices given by the potential $V$ in $\{x\geq
0\}$ and a rotated copy of $V$ for $\{x<0\}$; we
obtain an interface at $x=0$ where the two copies meet under the
defect angle $\theta$. Our main assumption is that the periodic
operator $H:=R_0$ has a non-trivial gap $(a,b)$.
 We then have $R_\theta\to R_{\theta_0}$ in the strong resolvent sense
 as $\theta\to\theta_0 \in [0,\pi/2)$; in particular $R_\theta$ converges to $H$ in the
strong resolvent sense as $\theta\to 0$. Our main result, Theorem 1.2 below,
 shows that the spectrum of $R_\theta$ is discontinuous at $\theta = 0$; in
 particular,  $R_\theta$ cannot converge to $H$ in the norm resolvent sense as
$\theta \to 0$.
\vskip1em

{\bf 1.2.~Theorem.} {\it Let $H$, $R_\theta$ and $(a,b)$ as above with
 a Lipschitz-continuous potential $V$.
Then, for any $\eps>0$, there exists $0<\theta_\eps<\pi/2$ such
that for any $E\in(a,b)$ we have
$$
\sigma(R_\theta)\cap(E-\eps,E+\eps)\neq\emptyset,\quad\forall
0<\theta<\theta_\eps.\eqno{(1.4)}
$$
}

As an illustration, we will consider potentials of {\it muffin tin
type} which can be specified by fixing a radius $0 < r < 1/2$ for
the discs where the potential vanishes, and the center $P_0 =
(x_0, y_0) \in [0,1)^2$ for the generic disc. In other words, we
consider the periodic sets
$$
\Omega_{r,P_0} := \cup_{(i,j) \in \Z^2} B_r(P_0 +
(i,j)),\eqno{(1.5)}
$$
and we let $V = V_{r,P_0}$ be zero on $\Omega_{r,P_0}$ while we
assume that $V$ is infinite on $\R^2 \setminus \Omega_{r,P_0}$. If
$H_{i,j}$ is the Dirichlet Laplacian of the disc $B_r(P_0 +
(i,j))$, then the form sum of $-\Delta$ and $V_{r,P_0}$ is
$\oplus_{(i,j)\in \Z^2} H_{i,j}$. In our examples, we can see the
behavior of surface states in the dislocation problem and the
rotation problem for $-\Delta+V_{r,P_0}$ directly.

The paper is organized as follows: Section 2 is devoted to the
dislocation problem on $\R$, on $\Sigma$ and in $\R^2$. Section 3
is about a small angle defect model in 2D and explains some
details of the proof of Theorem 1.2. Finally, in Section 4, we
turn to dislocations and rotations for muffin tin potentials
where results analogous to Theorem 1.1 and Theorem 1.2 can be obtained.
For further reading, we refer to [HK1, HK2].
%
%
\section{Dislocation problems on the real
line, on the strip $\R\times[0,1]$, and in the plane}
%
%
In this section, we study Schr\"odinger operators in one and two
dimensions where the potential is
obtained from a periodic potential by a coordinate shift on
$\{x<0\}$. We begin with a brief overview of the
one-dimensional dislocation problem. In a second step, we study
the dislocation problem on the strip $\Sigma=\R\times[0,1]$
which provides a connection between the dislocation
problems in one and two dimensions. Finally, we deal
with dislocations in $\R^2$. Some of the results obtained in
this section will be used in our treatment of
rotational defects in the following section.

Let $h_0$ denote the (unique) self-adjoint extension of
$-\frac{\text{d}^2}{\text{\d} x^2}$ defined on
$\Cci{\R}$. Our basic class of potentials is given by
$$
\PP : = \left\{ V \in L_{1, \text{loc}}(\R,\R) \mid
\forall x \in \R : \, V(x+1) = V(x) \right\}.\eqno{(2.1)}
$$
Potentials $V \in \PP$ belong to the class $L_{\text{1, \text{loc, unif}}}(\R)$
which coincides with the Kato-class on the real
line; in particular, any $V \in \P$ has relative form-bound zero
with respect to $h_0$ and thus the form sum $H$ of $h_0$ and $V
\in \P$ is well-defined (cf. [CFrKS]).

For $V \in \P$ and $t \in [0,1]$, we define the dislocation potentials
$W_t$ by $W_t(x) := V(x)$, for $x \ge 0$, and $W_t(x) := V(x+t)$, for $x <
0$.  As before, the form-sum $H_t$ of $h_0$ and $W_t$ is well-defined.

We begin with some well-known results pertaining to the spectrum
of $H = H_0$. As explained in [E, RS-IV], we have
$$
\sigma(H) = \sigmaess(H) = \cup_{k=1}^\infty [\gamma_k,\gamma'_k],
\eqno{(2.2)}
$$
where the numbers $\gamma_k$ and $\gamma'_k$ satisfy $\gamma_k < \gamma_k'
\le \gamma_{k+1}$, for all $k \in \N$, and $\gamma_k \to \infty$
as $k\to\infty$. Moreover, the spectrum of $H$ is purely
absolutely continuous. The intervals $[\gamma_k,\gamma_k']$ are
called the {\it spectral bands} of $H$. The open intervals
$\Gamma_k := (\gamma_k',\gamma_{k+1})$ are the {\it spectral gaps}
of $H$; we say the $k$-th gap is {\it open} or {\it
non-degenerate} if $\gamma_{k+1} > \gamma_k'$.
It is easy to see ([HK1]) that
$$
\sigmaess(H_t)=\sigmaess(H),\quad 0\leq t\leq 1,
\eqno{(2.3)}
$$
since inserting a Dirichlet boundary condition at a finite number of points
means a finite rank perturbation of the resolvent, as is well known. Hence each
non-trivial gap $(a,b)$ of $H$ is a gap in the essential spectrum of
$H_t$, for all $t$. However, the dislocation  may produce discrete (and simple)
eigenvalues inside the spectral gaps of $H$:
for any $(a,b)$ with $\inf\sigmaess(H)<a<b$ and $(a,b)\cap\sigma(H)=\emptyset$ there exists
$t\in(0,1)$ such that
$$
\sigma(H_t)\cap(a,b)\neq\emptyset.
\eqno{(2.4)}
$$
We thus have the following picture: while the essential spectrum remains
unchanged under the perturbation, eigenvalues of $H_t$ cross the (non-trivial)
gaps of $H$ as $t$ ranges through $(0,1)$. These eigenvalues of $H_t$
can be described by continuous functions of $t$ (cf.\ [K1, K2] and Lemma
2.1 below). Lemma 2.1 states the (more or less obvious) fact that
the eigenvalues of $H_t$ inside a given gap $\Gamma_k$ of $H$ can be described
by an (at most) countable, locally finite family of continuous functions,
defined on suitable subintervals of $[0,1]$. The proof of Lemma 2.1 uses a
straight-forward compactness argument (cf.\ [HK1]). The result stated in Lemma 2.1
is presumably far from optimal if one assumes periodicity
of the potential. On the other hand, the lemma and its proof in [HK1]
allow for a generalization to non-periodic situations.

\vskip1em
{\bf 2.1.~Lemma.} {\it Let $V\in\PP$ and $k \in \N$ and suppose that the gap
$\Gamma_k$ of $H$ is open. Then there is a (finite or countable)
family of continuous functions $f_j \colon (\alpha_j,\beta_j) \to
\Gamma_k$, where $0 \le \alpha_j < \beta_j \le 1$, with the
following properties:
\vskip.5ex
$(i)$ For all $j$ and for all $\alpha_j < t < \beta_j$,
 $f_j(t)$ is an eigenvalue of $H_t$. Conversely, for any $t \in (0,1)$ and
any eigenvalue $E \in \Gamma_k$ of $H_t$ there is a unique index
$j$ such that $f_j(t) = E$.
\vskip.5ex
$(ii)$ As $t \downarrow \alpha_j$ (or $t \uparrow \beta_j$), the
limit of $f_j(t)$ exists and belongs to the set $\{a,b\}$.
\vskip.5ex
$(iii)$ For all but a finite number of indices $j$ the range of
$f_j$ does not intersect a given compact subinterval of
$\Gamma_k$. }
\vskip1em
Under stronger assumptions on $V$ one can show that the eigenvalue
branches are H\"older- or Lipschitz-continuous, or even analytic
(cf.~[K1]): we consider potentials from the classes
$$
\PP_\alpha : = \left\{ V \in \PP\bigg|\; \exists C \ge 0 \colon
\int_0^1 |V(x + s) - V(x)| \d x \le C s^\alpha, \, \forall 0< s
\le 1\right\},\eqno{(2.5)}
$$
where $0 < \alpha \le 1$. The class $\PP_\alpha$ consists of all
periodic functions $V \in \P$ which are (locally)
$\alpha$-H\"older-continuous in the $L_1$-mean; for $\alpha = 1$
this is a Lipschitz-condition in the $L_1$-mean. The class $\PP_1$
is of particular practical importance since it contains the
periodic step functions. As shown by J.\ Voigt,  $\P_1$ coincides
with the class of periodic functions on the real line which are
locally of bounded variation (cf. [HK1]).
\vskip1em
{\bf 2.2.~Proposition.} {\it For $V \in \P_1$, let $(a,b)$ denote
any of the gaps $\Gamma_k$ of $H$ and let $f_j \colon (\alpha_j,
\beta_j) \to (a,b)$ be as in Lemma 2.1. Then the functions $f_j$
are uniformly Lipschitz-continuous. More precisely, there exists a
constant $C \ge 0$ such that for all $j$
$$
|f_j(t) - f_j(t')| \le C |t - t'|, \qquad \alpha_j \le t, t' \le
\beta_j.\eqno{(2.6)}
$$
If $0<\alpha <1$ and $V \in \P_\alpha$, then each of the functions
$f_j \colon (\alpha_j,\beta_j) \to (a,b)$ is locally uniformly
H\"older-continuous, i.e., for any compact subset $[\alpha_j',
\beta_j'] \subset (\alpha_j, \beta_j)$ there is a constant $C =
C(j, \alpha_j', \beta_j')$ such that $|f_j(t) - f_j(t')| \le C |t
- t'|^\alpha$, for all $t, t' \in [\alpha_j', \beta_j']$. }
\vskip1em
Our basic result in the study of the one-dimensional dislocation
problem says that at least $k$ eigenvalues move from the upper to
the lower edge of the $k$-th gap as the dislocation parameter
ranges from $0$ to $1$. Using the notation of Lemma 2.1 and
writing $f_i(\alpha_i) := \lim_{t\downarrow\alpha_i}f_i(t)$,
$f_i(\beta_i) := \lim_{t\uparrow\beta_i}f_i(t)$, we define
$$
\NN_k := \#\{ i \mid f_i(\alpha_i) = b, \,\,\, f_i(\beta_i) = a \}
-  \#\{ i \mid f_i(\alpha_i) = a, \,\,\, f_i(\beta_i) = b \}
\eqno{(2.7)}
$$
(note that both terms on the RHS of eqn.~(2.7) are finite by Lemma 2.1 ($iii$)).
Thus $\NN_k$ is precisely the number of eigenvalue branches of
$H_t$ that cross the $k$-th gap moving from the upper to the lower
edge minus the number crossing from the lower to the upper edge.
Put differently, $\NN_k$ is the spectral multiplicity which {\it
effectively} crosses the gap $\Gamma_k$  in downwards direction as
$t$ increases from $0$ to $1$. We then have the following result.
%
%
\vskip1.0em
{\bf 2.3.~Theorem.} (cf.\ [K1, HK1])

{\it Let $V \in \PP$ and let $k\in\N$ be such
that the $k$-th spectral gap of $H$ is open, i.e., $\gamma_k' <
\gamma_{k+1}$. Then $\NN_k = k$.}
\vskip1.0em
In fact, the results obtained by Korotyaev in [K1, K2] are more
detailed; e.g., Korotyaev shows that the dislocation operator produces
at most two states (an eigenvalue and a resonance) in a gap of the
periodic problem. On the other hand, our variational arguments are
more flexible and allow an extension to higher dimensions, as we
will see in the sequel. The main idea of our proof---somewhat
reminiscent of [DH, ADH]---goes as follows: consider a sequence
of approximations on intervals $(-n-t,n)$ with associated
operators $H_{n,t} = -\frac{\text{d}^2}{\text{d}x^2}+ W_t$ with
periodic boundary conditions. We first observe that the gap
$\Gamma_k$ is free of eigenvalues of $H_{n,0}$ and $H_{n,1}$ since
both operators are obtained by restricting a periodic operator on
the real line to some interval of length equal to an entire
multiple of the period, with periodic boundary conditions. Second,
the operators $H_{n,t}$ have purely discrete spectrum and it
follows from Floquet theory (cf. [E, RS-IV]) that $H_{n,0}$ has
precisely $2n$ eigenvalues in each band while $H_{n,1}$ has
precisely $2n+1$ eigenvalues in each band. As a consequence,
effectively $k$ eigenvalues of $H_{n,t}$ must cross any fixed
$E\in\Gamma_k$ as $t$ increases from $0$ to $1$. To obtain the result
of Theorem 2.3 we only have to take the limit $n \to \infty$; cf.\
[HK1] for the technical arguments. In [HK1], we also discuss a
one-dimensional periodic step potential and perform some explicit
(and also numerical) computations resulting in a plot of an
eigenvalue branch for the associated dislocation
problem.\vskip.5em

We now turn to the dislocation problem on the infinite strip
$\Sigma=\R \times[0,1]$. Let $V \colon \R^2 \to \R$
be $\Z^2$-periodic and Lipschitz continuous. We denote by $S_t$
the (self-adjoint) operator $-\Delta + W_t$, acting in
$L_2(\Sigma)$, with periodic boundary conditions in the
$y$-variable and with $W_t$
defined as in eqn.~(1.1); again, the parameter $t$ ranges between
$0$ and $1$. Since $S_0$ is periodic in the $x$-variable, its
spectrum has a band-gap structure. To see that the essential
spectrum of the family $S_t$ does not depend on the parameter $t$,
i.e., $\sigmaess(S_t)=\sigmaess(S_0)$ for all $t\in[0,1]$, it
suffices to prove compactness of the resolvent difference
$(S_t-c)^{-1}-(S_{t,D}-c)^{-1}$, where $S_{t,D}$ is $S_t$ with an
additional Dirichlet boundary condition at $x=0$, say. (While, in
one dimension, adding in a Dirichlet boundary condition at a
single point causes a rank-one perturbation of the resolvent, the
resolvent difference is now Hilbert-Schmidt, which can be seen
from the following well-known line of argument: If
$-\Delta_\Sigma$ denotes the (negative) Laplacian in $L_2(\Sigma)$
and $-\Delta_{\Sigma;D}$ is the (negative) Laplacian in
$L_2(\Sigma)$ with an additional Dirichlet boundary condition at
$x=0$, then $(-\Delta_\Sigma + 1)^{-1} - (-\Delta_{\Sigma;D} +
1)^{-1}$ has an integral kernel which can be written down
explicitly using the Green's function for $-\Delta_\Sigma$ and the
reflection principle.)

While the band gap structure of the essential spectrum of $S_t$
 is independent of $t \in [0,1]$, $S_t$ will have discrete
eigenvalues in the spectral gaps of $S_0$ for appropriate values
of $t$.
We have the following result.
\vskip1.0em

{\bf 2.4.~Theorem.} {\it Assume that $V$ is Lipschitz-continuous. Let $(a,b)$ denote a non-trivial spectral gap
of $S_0$ and let $E \in (a,b)$. Then there
exists $t = t_E \in (0,1)$ such that $E$ is a discrete eigenvalue
of $S_t$. }
\vskip1.0em
As on the real line, we work with approximating problems on finite
size sections of the infinite strip $\Sigma$. Let
$\Sigma_{n,t}:=(-n-t,n)\times(0,1)$ for $n\in\N$, and consider
$S_{n,t}:=-\Delta+W_t$ acting in $L_2(\Sigma_{n,t})$ with periodic
boundary conditions in both coordinates. The operator $S_{n,t}$
has compact resolvent and purely discrete spectrum accumulating
only at $+\infty$. The rectangles $\Sigma_{n,0}$ (respectively,
$\Sigma_{n,1}$) consist of $2n$ (respectively, $2n+1$) period
cells. By routine arguments (see, e.g., [RS-IV, E]), the number of
eigenvalues below the gap $(a,b)$ is an integer multiple of the
number of cells in these rectangles; we conclude that eigenvalues
of $S_{n,t}$ must cross the gap as $t$ increases from $0$ to $1$.
Thus for any $n\in\N$ we can find $t_n\in(0,1)$ such that
$E\in\sigmadisc(S_{n,t_n})$; furthermore, there are eigenfunctions $u_n\in
D(S_{n,t_n})$ satisfying $S_{n,t_n}u_n = Eu_n$, $\norm{u_n}=1$, and
$\norm{\nabla u_n} \le C$ for some constant $C \ge 0$. Multiplying
 $u_{n}$ with a  suitable cut-off function, we obtain (after
extracting a suitable subsequence) functions $v_{n}\in D(S_t)$ and
$t\in(0,1)$ satisfying
$$
\norm{(S_{t} - E) v_{n}} \to 0\quad\text{and}\quad\norm{v_{n}}
  \to 1, \eqno{(2.8)}
$$
as $n \to \infty$, which implies $E \in \sigma(S_{t})$, cf.
[HK1].\vskip.5em

Finally, we consider the dislocation problem on the plane $\R^2$
where we study the operators
$$
D_t=-\Delta+W_t,\quad 0\leq t\leq 1. \eqno{(2.9)}
$$
Denote by $S_t(\theta)$ the operator $S_t$ on the strip $\Sigma$
 with $\theta$-periodic boundary conditions in the $y$-variable.
 Since $W_t$ is periodic with respect to $y$,
 we have
$$
D_t\simeq\int_{[0,2\pi]}^{\oplus}S_t(\theta)\frac{\d\theta}{2\pi};
\eqno{(2.10)}
$$
in particular, $D_t$ has no singular continuous part, cf.\ [DS]. As
for the spectrum of $S_t$ inside the gaps of $S_0$, Theorem 2.4
yields the following result.
\vskip1em

{\bf 2.5.~Theorem.} {\it Assume that $V$ is Lipschitz-continuous. Let $(a,b)$ denote a non-trivial spectral gap of
$D_0$ and let $E \in (a,b)$. Then
there exists $t = t_E \in (0,1)$ with $E \in \sigma(D_t)$. }
\vskip1em

{\bf Proof.} Let $v_n\in D(S_t)$ denote an approximate solution of
the eigenvalue problem for $S_t$ and $E$; see (2.8).
We extend $v_n$ to a function ${\tilde v}_n(x,y)$ on
$\R^2$ which is periodic in $y$. By multiplying $\tilde v_n$ by
smooth cut-off functions $\Phi_n(x,y)$, we obtain functions
$$
w_n = w_n(x,y) := \frac{1}{\norm{\Phi_n\tilde v_n}} \Phi_n\tilde
v_n \eqno{(2.11)}
$$
belonging to the domain of $D_t$
and satisfying $\norm{w_n} =1$, $\supp w_n \subset [-n,n]^2$, and
$$
(D_t - E) w_n \to 0, \qquad n \to \infty; \eqno{(2.12)}
$$
this implies the desired result.
\hfill$\square$\\[.5cm]

The stronger statement in Theorem 1.1. follows by a very similar line of argument.
The upshot is that the dislocation moves enough states through the
gap to have a non-trivial (integrated) surface density of states, for suitable
parameters~$t$.

The lower estimate established in Theorem 1.1. is
complemented by an upper bound which is of the expected order (up
to a logarithmic factor) in [HK2]. Note that the situation treated
in [HK2] is far more general than the rotation or
dislocation problems studied so far. In fact, here we allow for
different potentials $V_1$ on the left and $V_2$ on the right
which are only linked by the assumption that there is a common
spectral gap; neither $V_1$ nor $V_2$ are required to be periodic.
The proof uses technology which is fairly standard and is based on
exponential decay estimates for resolvents, cf. [S].
\vskip1.5em
{\bf 2.6.~Theorem.} {\it Let $V_1$, $V_2 \in L_\infty(\R^2,\R)$
and suppose that the interval $(a,b) \subset \R$ does not
intersect the spectra of the self-adjoint operators $H_k :=
-\Delta + V_k$, $k = 1, 2$, both acting in the Hilbert space
$L_2(\R^2)$. Let
$$
W := \chi_{\{x<0\}} \cdot V_1 + \chi_{\{x \ge 0\}} \cdot V_2
\eqno{(2.13)}
$$
and define $H:=-\Delta+W$, a self-adjoint operator in $L_2(\R^2)$.
Finally, we let $H^{(n)}$ denote the self-adjoint operator
$-\Delta+W$ acting in $L_2(Q_n)$ with Dirichlet boundary
conditions.
Then, for any interval $[a',b'] \subset (a,b)$, we have
$$
\limsup_{n\to\infty}\frac{1}{n\log n}N_{[a',b']}(H^{(n)}) <
\infty, \eqno{(2.14)}
$$
where $N_{[a',b']}(H^{(n)})$ denotes the number of eigenvalues of
$H^{(n)}$ in $[a',b']$.
}
\vskip1em
We note that the factor  $\log n$ in eqn.~(2.14) can presumably be
dropped under appropriate assumptions (H.~Cornean, private communication);
however, this seems to require substantially different, and less elementary, methods.
%
%
\section{Rotational defect in a
two-dimensional lattice}
%
%
In this section, we will use our results on the translational
problem to obtain spectral information about rotational problems
in the limit of small angles. Our main theorem deals with the
following situation. Let $V \colon \R^2 \to \R$ be a
Lipschitz-continuous function which is periodic w.r.t.\ the
lattice $\Z^2$. For $\theta \in (0,\pi/2)$, let
$$
M_\theta := \left(
\begin{array}{cc}
\cos\theta & -\sin\theta \\
\sin\theta & \cos\theta \\
\end{array}\right) \in \R^{2 \times 2},
\eqno{(3.1)}
$$
and $V_\theta$ as in (1.3). We then let $H_0$ denote the (unique)
self-adjoint extension of $-\Delta \upharpoonright{\Cci{\R^2}}$,
acting in the Hilbert space $L_2(\R^2)$, and
$$
R_\theta := H_0 + V_\theta, \qquad D(R_\theta) = D(H_0).
\eqno{(3.2)}
$$
Then $R_\theta$ is essentially self-adjoint on $\Cci{\R^2}$ and
semi-bounded from below.

Now our key observation consists in the following: for any $t \in (0,1)$
given, any $\eps > 0$, and any $n \in \N$, we can find points
$(0,\eta)$ on the $y$-axis with $\eta \in \N$ such that
$$
|V_\theta(x,y) - W_t(x,y)| < \eps, \qquad (x,y) \in Q_n(0,\eta),
\eqno{(3.3)}
$$
where $Q_n(0,\eta) = (-n,n) \times (\eta-n, \eta+n)$, provided
$\theta > 0$ is small enough and satisfies a condition which
ensures an appropriate alignment of the period cells on the
$y$-axis. Put differently: for very small angles, the
rotated potential $V_\theta$ will almost look like a dislocation
potential $W_t$, on suitable squares $Q_n(0,\eta)$.

To prove Theorem~1.2 we proceed as follows: Fix an arbitrary $E$
in a gap of $H = H_0 + V = R_0$. Knowing that there exists $t\in(0,1)$ such that
$E\in\sigmadisc(D_t)$, we choose an associated approximate
eigenfunction as constructed in the proof of Theorem~2.5 and
shift it along the $y$ axis until its support is contained in
$Q_n(0,\eta)$. In this way we obtain an approximate eigenfunction for
$R_\theta$ and $E$. It is easy to see that, in view of the
Lipschitz continuity and the $\Z^2$-periodicity of $V$, the
geometric conditions for the estimate (3.3) are
$$
|k\tan\theta-[k\tan\theta]-t|<\eps,\quad
|k/\cos\theta-\eta|<\eps 
\eqno{(3.4)}$$
for some $k \in \N$.
It thus remains to prove the existence of natural numbers $k$
satisfying the conditions in (3.4), for given $\theta\in(0,\pi/2)$.
Actually, the existence of numbers $k$ as desired can only be
established for a dense set of angles.
%
\vskip.5em
{\bf 3.1.~Lemma.} {\it Let $\T^2 = \R^2 / \Z^2$ be the flat
two-dimensional torus and let $T_\theta \colon \T^2 \to \T^2$ be
defined by
$$
T_\theta(x,y):=(x+\tan\theta,y+1/\cos\theta). \eqno{(3.5)}
$$
Then there is a set $\Theta \subset (0,\pi/2)$ with countable
complement such that the transformation $T_\theta$ in $(3.5)$ is
ergodic for all $\theta \in \Theta$.} \vskip1em
%
The assertion of Theorem~1.2 now follows from Birkhoff's ergodic theorem, cf.~[CFS, HK2]:
Let us first assume that $\theta\in\Theta$. Let $\eps>0$ and let us denote by $\chi=\chi_Q$
the characteristic function of the set
$Q:=(t-\eps,t+\eps)\times(-\eps,\eps)\subset\T^2$. Then
$$
\lim_{n\to\infty}\frac{1}{n}\sum_{m=0}^{n-1}\chi(T_\theta^m(0,0))=\int_{Q}\dx\d y=4\eps^2>0.
\eqno{(3.6)}
$$
By a simple approximation argument the statement of Theorem~1.2 also holds for angles $\theta\notin\Theta$.
Altogether, this completes the proof of Theorem 1.2.

Recall that strong resolvent convergence implies upper
semi-continuity of the spectrum while the spectrum may contract
considerably when the limit is reached. In the present section, we
are dealing with a situation where the spectrum in fact behaves
discontinuously at $\theta = 0$ since, counter to first intuition,
the spectrum of $R_\theta$ ``fills'' the gap $(a,b)$ as $ 0 \ne \theta
\to 0$. This implies, in particular, that
$R_\theta$ cannot converge to $H$ in the norm resolvent sense, as
$\theta \to 0$.
%
%
\section{Muffin Tin Potentials}
%
%
In this section, we present a class of examples where one can
arrive at rather precise statements that illustrate some of the
phenomena described before. Our potentials $V=V_{r,P_0}$ are of
muffin tin type, as defined in the introduction.
\vskip.5em

{\bf (1) The dislocation problem.} In the simplest case we would
take $x_0 = 1/2$ and $y_0 = 0$ so that the disks $B_r(1/2 + i, j)$,
for $i, j \in \Z$, will not intersect or touch the interface
$\{(x,y) \mid x=0\}$, for $0 < r < 1/2$. Defining the dislocation potential $W_t$ as
in (1.1), we see that there are bulk states given by the Dirichlet
eigenvalues of all the discs that do not meet the interface, and
there may be surface states given as the Dirichlet eigenvalues of
the sets $B_r(1/2 - t, j) \cap \{ x < 0\}$ for $j \in \Z$ and $1/2
- r < t < 1/2 + r $.

More precisely, let $\mu_k = \mu_k(r)$ denote the Dirichlet
eigenvalues of the Laplacian on the disc of radius $r$, ordered by
min-max and repeated according to their respective multiplicities.
The Dirichlet eigenvalues of the domains $B_r(1/2 - t, 0) \cap \{x
< 0\}$, for $1/2 - r < t < 1/2 + r$, are denoted as $\lambda_k(t) =
\lambda_k(t,r)$; they are continuous, monotonically decreasing
functions of $t$ and converge to $\mu_k$ as $t \uparrow 1/2 + r$
and to $+\infty$ as $t \downarrow 1/2 - r$. In this simple model,
the eigenvalues $\mu_k$ correspond to the bands of a periodic
operator. We see that the gaps are crossed by surface states as
$t$ increases from $0$ to $1$, in agreement with Theorem 1.1. In
[HK1] we also discuss muffin tin potentials with dislocation in
the $y$ direction.
\vskip.5em
{\bf (2) The rotation problem.} In [HK2], we look at three types
of muffin tin potentials and discuss the effect of the ``filling
up'' of the gaps at small angles of rotation. We begin with muffin
tins with walls of infinite height, then approximate by muffin tin
potentials of height $n$, for $n \in \N$ large. By another
approximation step, one may obtain examples with
Lipschitz-continuous potentials. These examples show, among other
things, that Schr\"odinger operators of the form $R_\theta$ may in
fact have spectral gaps
for some $\theta > 0$. For the sake of brevity, we
only state our main results and refer to [HK2] for further
details.

We write (in the notation of (1.5)) $\Omega_r=\Omega_{r,(1/2,1/2)}$ and
$$
\Omega_{r,\theta} := \Omega_r \cap \{ x \ge 0\} \cup (M_\theta
\Omega_r) \cap \{x < 0 \}, \eqno{(4.1)}
$$
and let $H_{r,\theta}$ denote the Dirichlet Laplacian on
$\Omega_{r,\theta}$ for $0<r<1/2$ and $0\leq\theta\leq\pi/4$.
Denote the Dirichlet eigenvalues of the Laplacian $H_r$ of
$\Omega_r$ by $(\tilde\mu_j(r))_{j\in\N}$, with $\tilde\mu_j(r)\to\infty$ as
$j\to\infty$ and $\tilde\mu_j(r)<\tilde\mu_{j+1}(r)$ for all $j\in\N$; note
that the eigenvalues $\tilde\mu_j$ may have multiplicity $> 1$.

\vskip1em
{\bf 4.1.~Proposition.} {\it
Let $(a,b)$ be one of the gaps $(\tilde\mu_j,\tilde\mu_{j+1})$
 and let $0 < r < 1/2$ be fixed.

$(a)$ Each $\tilde\mu_j(r)$, $j = 1, 2, \ldots$, is an eigenvalue of
infinite multiplicity of $H_{r,\theta}$, for all $0 \le \theta
\le\pi/2$. The spectrum of $H_{r,\theta}$ is pure point, for all
$0 \le \theta \le \pi/2$.

$(b)$ For any $\eps > 0$ there is a $\theta_\eps = \theta_\eps(r)
> 0$ such that any interval $(\alpha, \beta) \subset (a,b)$ with
$\beta - \alpha \ge \eps$ contains an eigenvalue of $H_{r,\theta}$
for any $0 < \theta < \theta_\eps$.

$(c)$ There exists a set $\Theta \subset (0,\pi/2)$ of full
measure such that $\sigma(H_{r,\theta}) = [{\tilde\mu}_1(r),\infty)$. The
eigenvalues different from the ${\tilde\mu}_j(r)$ are of finite
multiplicity for $\theta \in \Theta$.} \vskip1em
{\bf 4.2.~Remark.}
 If $\tan\theta$ is rational, the grid $M_\theta \Z^2$
 is periodic in the $x$- and $y$-directions with $\theta$-dependent periods
 $p, q \in \N$. As a consequence, $H_{r,\theta}$ has at most a finite number
 of eigenvalues in $(a,b)$ for $\tan\theta$ rational, each of them of infinite multiplicity.
 Hence we see a  drastic change in the spectrum for $\tan\theta \in \Q$ as
 compared with $\theta \in \Theta$.
 Furthermore, if $\tan\theta$ is rational with $\tan\theta \notin \{ 1/(2k+1) \mid k \in
 \N \}$, then there is some $r_\theta > 0$ such that $\sigma(H_{r,\theta}) =
 \sigma(H_r)$ for all $0 < r < r_\theta$.

\vskip1em
We next turn to muffin tin potentials of finite height. Here we
define the potential $V_{r,\theta}$ to be zero on
$\Omega_{r,\theta}$ and $V_{r,\theta} = 1$ on the complement of
$\Omega_{r,\theta}$, where $0 < r < 1/2$ and $0 \le \theta \le
\pi/4$; we also let $H_{r,n,\theta} := H_0 + n V_{r,\theta}$.  The
periodic operators $H_{r,n,0}$ have purely absolutely continuous
spectrum and $H_{r,n,\theta}\to H_{r,\theta}$ in the sense of norm
resolvent convergence, uniformly for $\theta\in[0,\pi/4]$.
\vskip1.5em
{\bf 4.3.~Proposition.} {\it Let $(a,b)$ be one of the gaps
  $(\tilde\mu_j,\tilde\mu_{j+1})$. We then have:

$(a)$ For $\tan\theta \in \Q$ the spectrum of $H_{r,n,\theta}$ has
gaps inside the interval $(a,b)$ for $n$ large. More precisely,
if $H_{r,\theta}$ has a gap $(a',b') \subset (a,b)$,
then, for $\eps> 0$ given,  the interval $(a'+\eps, b'-\eps)$ will be free of
spectrum of $H_{r,n,\theta}$ for $n$ large.

$(b)$ For any $\eps > 0$ there are $\theta_0 > 0$ and $n_0 > 0$
such that any interval $(c-\eps, c+\eps) \subset (a,b)$ contains
spectrum of $H_{r,n,\theta}$ for all $0<\theta<\theta_0$ and $n
\ge n_0$.

}
\vskip1em
By similar arguments, we can approximate $V_{r,\theta}$ by
Lip\-schitz-continuous muffin tin potentials that converge
monotonically (from below) to $V_{r,\theta}$ in such a way that
norm resolvent convergence holds for the associated Schr\"odinger
operators (again uniformly in $\theta \in [0,\pi/4]$). The
spectral properties obtained are analogous to the ones stated in
Proposition 4.3. Note, however, that the statement corresponding
to part $(b)$ in Proposition 4.3 is weaker than the result of our
main Theorem 1.1.
%
%
%
\subsection*{Acknowledgements.} The authors thank E.\ Korotyaev (St.\ Petersburg)
and J.\ Voigt (Dresden) for useful discussions.
\end{document}